\documentstyle[preprint,aps,psfig]{revtex}
\begin{document}

\title{Mechanism of Poisoning the Catalytic Activity\\
 of Pd(100) by a Sulfur Adlayer}

\author{S. Wilke\cite{prad} and M. Scheffler}
\address{Fritz-Haber-Institut der Max-Planck-Gesellschaft,
Faradayweg 4-6, D-14\,195 Berlin--Dahlem, Germany}

\date{\today}

\maketitle

\begin{abstract}

The modification of the potential-energy surface (PES) of H$_2$ dissociation
over Pd(100) as induced by the presence of a (2$\times$2)\ S adlayer is
investigated by density-functional theory and the linear augmented plane wave
method. It is shown that the poisoning effect of S originates from the
formation of energy barriers hindering the dissociation of H$_{\text{2}}$. The
barriers are in the entrance channel of the PES and their magnitude strongly
depends on the lateral distance of the H$_2$ molecule from the S adatoms.

\end{abstract}


The modification of the chemical reactivity and selectivity of metal surfaces
by adsorbates is a most important ingredient of catalysis.  Nevertheless, the
understanding of the microscopic mechanism of how adatoms promote or poison
certain reactions is grossly incomplete.  Hydrogen dissociation has become the
benchmark system for theoretical and experimental studies of a simple chemical
reaction, and much progress has been made in developing a picture of how H--H
bonds are broken on clean metal surfaces, and how new bonds between the
hydrogen atoms and the surface are
formed~\cite{lun79,nor81,feib91,ham94,whit94,ham95,whit95,wil95b}. However, a
quantitative study of the modifications of the potential-energy surface (PES)
of hydrogen dissociation on an adlayer-covered surface has not been addressed
by theory so far.

The Pd(100) surface is probably the best-suited substrate for an investigation
of the mechanism of poisoning a catalytic reaction, because many experimental
and theoretical studies exist and form a good body of information to compare
to. At the clean Pd(100) surface hydrogen dissociates spontaneously,
i.e. dissociation pathways exist which have no hampering energy
barrier~\cite{wil95b,com80,ren88,bur90,behm80}.  However, when the surface is
covered with sulfur the H$_2$ sticking probability is reduced by several
orders of magnitude~\cite{com80,ren88,bur90}. The mechanism actuating this
huge poisoning effect is not known. Using density functional theory
calculations we show in this Letter that adsorbed sulfur builds up energy
barriers which hinder the dissociation, and that, because these energy
barriers are located in the entrance channel of the dissociation-reaction
pathway, this hinderance is particularly effective.

The influence of adsorbed sulfur on the dissociative adsorption of
H$_{\text{2}}$ has been studied using molecular beam experiments~\cite{ren88},
temperature programmed desorption (TPD) of hydrogen molecules adsorbed at
$T=110$ K~\cite{bur90}, and energy resolved studies of molecules desorbing
from the surface after penetrating a thin Pd film~\cite{com80}. The results
show that with increasing sulfur coverage $\Theta_{\text{S}}$ the initial
sticking coefficient of H$_2$ strongly decreases~\cite{ren88,bur90}. This is
true in particular for molecules with low kinetic energy~\cite{ren88}
($\lesssim$ 0.1 eV), as these can adsorb only if non-activated dissociation
pathways exist and are accessible. For $\Theta_{\text{S}}$=0.25 the initial
sticking coefficient of molecules with low energies is approximately two
orders of magnitude smaller than that for the clean surface~\cite{ren88,bur90}
indicating that for this sulfur coverage non-activated dissociation is nearly
completely hindered. When the sulfur coverage is increased the H$_2$ sticking
coefficient reduces even further (at $\Theta_{\text{S}}=0.5$ it is about three
orders of magnitude smaller than at the clean surface). In addition to this
dramatic decrease of the initial sticking coefficient TPD studies~\cite{bur90}
observed a decrease of the hydrogen saturation coverage with increasing
$\Theta_{\text{S}}$. Burke and Madix~\cite{bur90} therefore concluded that
sulfur adatoms substantially reduce the hydrogen adsorption energy at sites in
their vicinity making these positions unstable against associative
desorption. This was also the conclusion of earlier permeation studies of
Comsa et al.~\cite{com80}.

Theoretical studies have addressed this problem as well. Feibelman and
Hamann~\cite{feib84a,feib84b} suggested that the poisoning effect of sulfur is
related to the S induced change of the density of states (DOS) at the Fermi
level (see also Ref.~\cite{lar86}).  A different mechanism which could give
rise to a modification of the reactivity by adlayers is the interaction of the
H$_2$ molecule with the adlayer induced electrostatic
field~\cite{nor93,ham93b}. In all this work it remained open, how the
described effect will modify the hydrogen sticking (by blocking adsoprtion
sites for atomic hydrogen or by building up energy barriers along the
dissociation-reaction pathway of H$_2$).

In a previous paper~\cite{wil95a} we reported density functional theory
calculations of the adsorption energy and geo\-metry of hydrogen at different
adsorption sites on clean Pd(100) and on Pd(100) covered with a (2$\times$2)\
S adlayer ($\Theta_{\text{S}}$=0.25). The calculations show that the presence
of a (2$\times$2)\ S adlayer moderately decreases the hydrogen adsorption
energy, but the reduction is not sufficient for a strict blocking of hydrogen
adsorption sites~\cite{wil95a}.  The poisoning effect of S is not governed by
a decrease of the adsorption energy alone and a complete understanding of
poisoning by adsorbed sulfur requires a calculation of the PES for hydrogen
dissociation in the presence of a sulfur adlayer. In the present paper we
report and discuss such calculations. They were performed using
density-functional theory with the generalized gradient approximation
(GGA)~\cite{per92} for the exchange-correlation functional. The full potential
linear augmented plane wave (FP-LAPW) method~\cite{wien93,fhiv} is employed
for solving the non-relativistic Kohn-Sham equations.  In the interstitial
region the wave functions are represented by a plane wave expansion up to
$E_{\text{cut}}$=11 Ry, and plane waves up to $\tilde{E}_{\text{cut}}$=169\ Ry
are taken into account for the potential representation. For the {\bf
k}-integration we used 32 uniformly-spaced points in the two-dimensional
Brillouin zone of the (2$\times$2) surface unit cell.  We used a supercell
geo\-metry and modeled the metal substrate by three-layer slabs which are
separated by a 10~{\AA} thick vacuum region. The adsorbates are placed at both
sides of the slab. The geo\-metry of the Pd(100) (2$\times$2) S surface is
determined by total-energy minimization~\cite{wil95a}. This geo\-metry is kept
fixed when the H$_2$ dissociation pathway is studied, which is the appropriate
and plausible approach, because the mass mismatch of the different atoms is
significant. The most severe approximations in our studies are the thinness of
the metal slab. For studies of adsorption energies an accurate treatment of a
fcc (100) surface typically requires to take at least five or seven layers
into account, when an accuracy of adsorption energies of about 0.1 eV is aimed
for.  In the present study of hydrogen dissociation we encounter the more
fortunate situation, that the PES, which is presented in Fig. 2, and which
corresponds to a three-layer Pd slab, is found to be indeed meaningful. This
is concluded from calculations at the geo\-metries of the energy barriers and
adsorption sites of Fig.~\ref{pes} using a five- and a seven layer slab (see
Table~1). The resulting energies and forces show that the topology of the PES
shown in Fig. \ref{pes} is correct and that the energy barriers change by less
than 0.05 eV when a thicker Pd slab is used.  We believe that the high number
of valence electrons of Pd and the fact that the Fermi level cuts the $d$ band
such that the possibility for low energy electron-hole excitations still
exists, are responsible for the fortunate situation that the PES of hydrogen
dissociation on a Pd crystal is well described by that at thin Pd slabs.

\begin{figure}
\psfig{figure=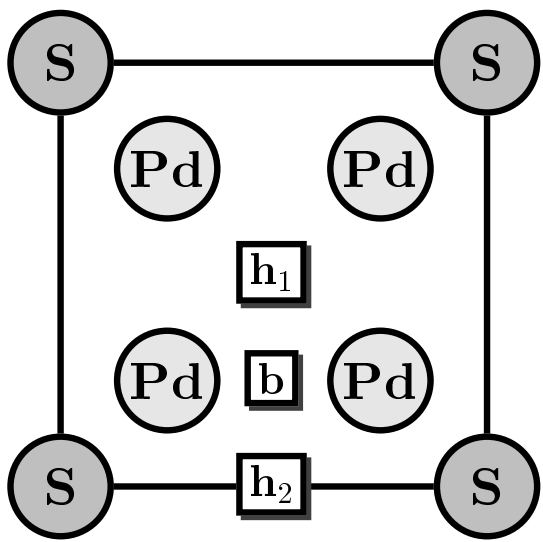,width=8cm}
\caption{Surface geo\-metry of (2$\times$2)S on Pd(100) with the two different
hollow sites labeled as ${\bf h_{1}}$ and ${\bf h_{2}}$ and the bridge site
${\bf b}$.}
\label{sketch}
\end{figure}

\begin{figure}
\psfig{figure=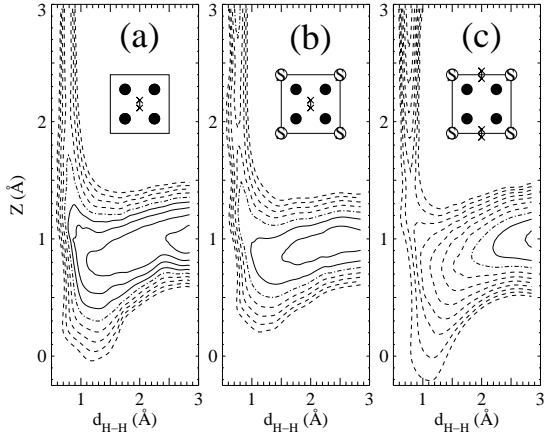,angle=90,width=10cm}
\caption{Cuts through the six-dimensional potential-energy surface (PES) of
H$_2$ dissociation over the surface hollow site of the clean Pd(100) surface
(a) and over the hollow sites ${\bf h_{1}}$ and ${\bf h_{2}}$ (c) at
(2$\times$2)S/Pd(100). The energy contours are displayed as a function of the
H--H distance, $d_{\text{H-H}}$, and the height of the H-H center of mass
above the center of the top Pd layer, $Z$.  The geo\-metry of each
dissociation pathway is indicated in the inset.  Positive values of the
potential energy are given by dashed, negative ones by solid lines. The zero
of energy is given as a dot-dashed line.  The interval between adjacent
contours is 0.1 eV.}
\label{pes}
\end{figure}

Figure~\ref{sketch} shows the surface unit cell for a sulfur coverage
$\Theta_{\text{S}}$=0.25. Out of the three hollow sites per unit cell only two
are distinct. They are denoted ${\bf h_{1}}$ and ${\bf h_{2}}$. The PES for
the H$_2$ dissociation at a rigid substrate is six dimensional.  However, the
influence of the (2$\times$2) sulfur adlayer on H$_2$ dissociation on Pd(100)
is well presented by three decisive cuts through the PES (see Fig.~\ref{pes}).
These, so-called elbow-plots, are defined as follows. The center-of-mass of
the two hydrogen nuclei is fixed over a surface hollow site, and the H--H axis
is kept parallel to surface, which is the energetically most favorable
orientation. Then the energy is displayed as a function of the H--H
interatomic distance and the height of the H--H center of mass above the
surface. Figure~\ref{pes}a shows the results for the dissociation over the
hollow site at the clean surface, and Figs.~\ref{pes}b,c are for the
(2$\times$2)S/Pd(100) substrate at the hollow site ${\bf h_{1}}$ and the
hollow site ${\bf h_{2}}$, respectively. The different geo\-metries of these
different cuts through the PES are given in the insets. The adsorption height
of sulfur adatoms on the (2$\times$2)\ S covered Pd(100) surface is calculated
as $Z_{\text{S}}$=1.24~{\AA} above the top Pd layer~\cite{wil95a}.  The PES of
H$_2$ dissociation on clean Pd(100) in Fig.~\ref{pes}a gives an example of
non-activated dissociation, where the two hydrogen nuclei may follow an always
downhill dissociation pathway~\cite{wil95b}. The presence of a (2$\times$2)\ S
adlayer on Pd(100) ($\Theta_{\text{S}}$=0.25) changes the dissociation process
significantly. The lowest-energy dissociation pathways at the sulfur-covered
surface in Fig.~\ref{pes}b,c have energy barriers of 0.1 eV and 0.6 eV. In
addition to the dissociation over the hollow sites (Fig.~\ref{pes}b,c), we
also considered the dissociation geo\-metry where the center-of-mass of the
molecule is situated over the bridge site ${\bf b}$ and the two H atoms point
towards the hollow sites ${\bf h_{1}}$ and ${\bf h_{2}}$. At the clean surface
this dissociation pathway is also non-activated and the decrease of the
potential energy along the dissociation pathway is even steeper than for
dissociation over the hollow site~\cite{wil95b}. We found that upon sulfur
adsorption the dissociation of H$_2$ becomes an activated process in this
geo\-metry too (the energy barrier is 0.15 eV). In addition to the formation
of energy barriers for different adsorption pathways the sulfur adlayer
reduces the adsorption energies for atomic hydrogen (see Fig. 2 at larger
$d_{\text{H-H}}$ values). The latter result was discussed in detail in
Ref.~\cite{wil95a}.  The PES shown in Fig.~\ref{pes} affirms that the
poisoning effect of sulfur adatoms for hydrogen dissociation on Pd(100) are
determined by the sulfur-induced formation of energy barriers. It is evident
and in fact plausible that the height of the barrier strongly depends on the
lateral distance of the hydrogen molecule from the sulfur adatoms.

At a first glance it is astonishing that such small energy barriers, which are
induced by the sulfur adsorption, can have the pronounced effect on the
dissociative adsorption of H$_{\text{2}}$ which was alluded to in the
introduction. However, the important theoretical result is not just the
heights of the barriers, but also their location.  The theory predicts that
the energy barriers, which are induced by S adsorption, are in the entrance
channel at a height above the top Pd layer of $Z$ = 1.8 and 1.4 {\AA}, and
$d_{\rm H-H} = 0.78$ {\AA}.  Thus, we encounter a different situation to that
found, e.g. for clean Cu and NiAl surfaces~\cite{ham94,whit94,ham95}. The
location of the energy barrier in the entrance channel implies (see
e.g. Ref.~\cite{hal90}) that for the impinging H$_2$ molecule the H--H bond
remains practically intact up to the position of the top of the barrier.  As a
consequence, vibrational and translational degrees of freedom of the H$_2$
molecule couple only weakly and the vibrational energy of the impinging H$_2$
cannot be utilized to overcome the barrier. For sulfur coverages up to
$\Theta_{\text{S}}$=0.25 we therefore predict that the sticking coefficient
shows no strong dependence on the H$_2$ vibrational state.  We also note that
the energy barrier in the entrance channel implies a significant reduction of
sticking of low energy (thermal) molecules.  The investigations of H$_2$
dissociation at clean Pd(100)~\cite{wil95b,gross95} had shown that the large
sticking coefficient of H$_2$ molecules with energy below about 0.2 eV at
Pd(100) results in part from the steering of molecules with initially
unfavorable orientations of the molecular axis towards the non-activated
dissociation pathways.  In the case that energy barriers are induced along
those pathways, the phase space of initial conditions that result in a
dissociation of low energy hydrogen molecules, and hence the sticking
coefficient, shrinks drastically, because in addition to the hindered
dissociation along the particular pathway the steering mechanism
\cite{gross95} is also disabled.

In order to identify the microscopic origin of the poisoning effect of sulfur
in more detail we analyze the surface electronic structure. At first we recall
that the trends of the reactivity of clean metal surfaces for the dissociative
adsorption of hydrogen molecules could be explained within a simple chemical
bonding model~\cite{ham95}. At noble and transition metal surfaces the
presence and magnitude of a dissociation barrier is mainly governed by the
direct interaction of the closed-shell orbital structure of the H$_2$ molecule
with the metal $d$ states.  This interaction is attractive as long as the
bonding states of the H$_2$ $\sigma_{\text{g}}$ and the H$_2$
$\sigma_{\text{u}}$ levels with the metal $d$~band dominate the
energetics. The interaction becomes repulsive, i.e. an energy barrier is built
up, when the corresponding antibonding states become occupied by a critical
amount. These antibonding states have their main weight just above the $d$
band edge.  Thus, for a true transition metal these antibonding states are
practically empty, and a small or vanishing energy barrier
results~\cite{ham95,wil95a,wil95b}.

\begin{figure}
\psfig{figure=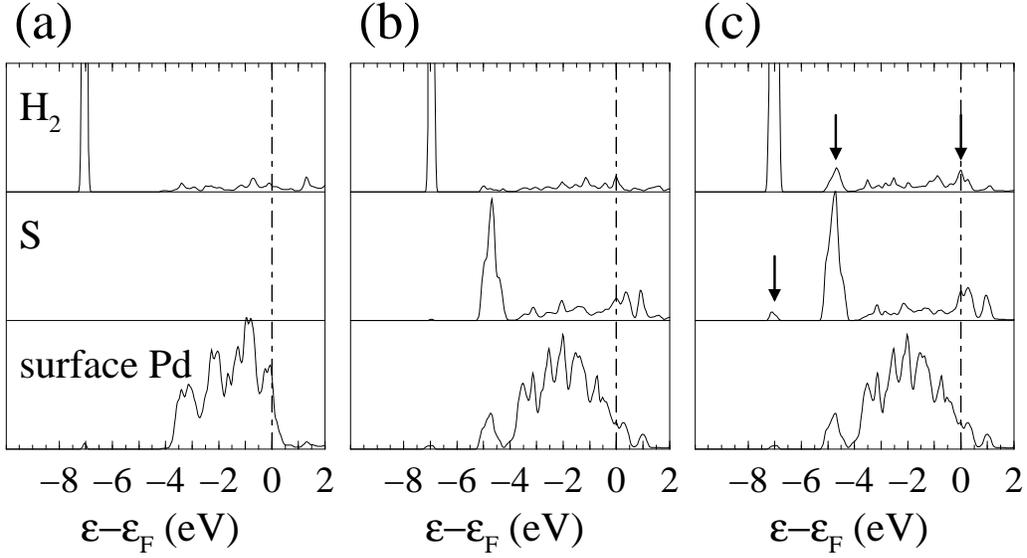,angle=270,width=14cm}
\vspace{5mm}

\caption{Density of states (DOS) for a H$_2$ molecule situated over a hollow
site at clean Pd(100) (a) and over the hollow sites ${\bf h_{1}}$ (b) and
${\bf h_{2}}$ (c) at (2$\times$2)\ S covered Pd(100). The Figure corresponds
to a particular point ($Z$=1.6 {\AA}, $d_{\text{H-H}}$=0.79 {\AA}) along the
reaction pathways shown in Fig.~2. Given is the local DOS at the H atoms, the
S adatom, and the surface Pd atoms within the respective muffin-tin sphere.}
\label{dos}
\end{figure}

In Fig.~\ref{dos} the densities of states (DOS) at three different lateral
positions of the H$_2$ molecule are compared; the height above the surface and
the H--H distance is taken to be the same ($Z$=1.6 {\AA},
$d_{\text{H-H}}$=0.79 {\AA}). It is evident that the sulfur $p$ orbitals
strongly interact with the Pd $d$ states, giving raise to a narrow peak just
below the Pd $d$ band edge (at $\epsilon - \epsilon_{\text{F}} = -4.8$~eV) and
a broad band at higher energies, which has substantial DOS at the Fermi level
(see middle panels of Figs. \ref{dos}b and c). The $d$ band at the surface Pd
atoms is considerably broadened due to the interaction with the S atoms (see
lower panels of Figs. \ref{dos} b and c).  The density of $d$ states close to
the Fermi energy is reduced by $\approx$50~\% compared to clean Pd(100) but
the value of the DOS at $\epsilon_{\text{F}}$ remains substantial.

The formation of a small energy barrier for hydrogen dissociation over the
${\bf h_{1}}$ hollow site due to the presence of the (2$\times$2)\ S adlayer
can be understood by extending the reactivity model discussed above. A
comparison of the DOS at the H atoms in Fig.~\ref{dos}a and b reveals that at
the sulfur-covered Pd surface the interaction of the $\sigma_{\text{g}}$
orbital of the H$_2$ molecule with the broadened band of the surface Pd $d$
states results in a broader distribution of states with an increased weight
below the Fermi level.  Consequently we encounter a larger occupation of
H$_2$--substrate antibonding states.  Thus, the repulsive contribution to the
H$_2$--surface interaction is increased and thus may give rise to the
formation of an energy barrier.  A H$_2$ molecule hitting the surface over the
hollow site ${\bf h_{1}}$ or ${\bf h_{2}}$ interacts with the same surface Pd
atoms and, hence, the formation of a large barrier over the hollow site ${\bf
h_{2}}$ cannot be explained by sulfur-induced modification of the electronic
structure at surface Pd atoms.  Intense peaks of the DOS at the H atoms for
the configuration where the H$_2$ molecule dissociates over the hollow site
${\bf h_{2}}$ (Fig.~\ref{dos}c) are found at energies of the sulfur related
bonding state at $-4.8$~eV and at the Fermi level. In addition there is a the
significant contribution of the sulfur adatom to the bonding states of the
H$_2$ $\sigma_{\text{g}}$--surface interaction at $-7$\ eV (see arrows in
Fig.~\ref{dos}c). These features clearly indicate that the H$_2$ molecule
interacts {\em directly} with states localized at the sulfur adatom.  We
conclude that the large energy barrier for H$_2$ dissociation over the hollow
site ${\bf h_{2}}$ is caused mainly by the repulsion between the occupied
H$_2$ $\sigma_{\text{g}}$ state and the spatially localized S--Pd bonding band
at $-4.8$~eV.  Direct H$_2$--S interactions become important
because due to the large adsorption height 
of the S adatoms H--S and H--Pd distances are comparable for larger heights of
the H$_2$ molecule in front of the surface. Indeed, as seen in Fig.~\ref{pes}c
the top of the energy barrier is situated at values of $Z$ close to the sulfur
adsorption height.

In conclusion, our DFT calculations of the modification of the PES of H$_2$
dissociation on Pd(100) due to the presence of a (2$\times$2) sulfur adlayer
show that the poisoning effect of sulfur adatoms is governed by the formation
of energy barriers which are situated in the entrance channel of the PES.  We
identified the direct interaction of the H$_2$ molecule with states at the S
adatom as the origin of the large barriers found over hollow sites neighboring
the S adatoms; at larger center-of-mass distances to the S adatom the
poisoning effect is weaker and results from the modification of the local
electronic structure at the surface Pd atoms.

\begin{table}
\begin{tabular}{ld|c|c|d}
        &   Fig. 2a & Fig.2b & Fig. 2c & Fig. 2b,c \\
  slab  &  $E_{\text{ad}}$ &  $E_{\text{b}}$ $\mid$ ($F_{{d}}$,
$F_{{Z}}$) &  $E_{\text{b}}$ $\mid$ ($F_{{d}}$,
$F_{{Z}}$)  &  $E_{\text{ad}}$  \\
  \tableline
      3     &      -0.47  &     0.09 $\mid$ (-66,-38) 
&      0.56 $\mid$ (-49,-41)  &    -0.21   \\
      5     &      -0.52  &     0.10 $\mid$ (-97,-54) 
&      0.59 $\mid$ (-72,-69)  &    -0.22  \\
      7     &      -0.54  &     0.09 $\mid$ (-90,-64) 
&      0.59 $\mid$ (-33,-90)  &    -0.22  \\
\end{tabular}
\caption{Dependence of the barrier heights, $E_{\rm b}$, and adsorption energy
at bridge sites, $E_{\rm ad}$, on the thickness of the substrate Pd slab.  The
positions refer to those of the three-layer calculation as given in Figs. 2a,
2b, and 2c. Units are in eV per H$_2$ molecule.  Included are the calculated
forces between between the hydrogen atoms $F_{d}$ and between the H$_2$
molecule and the surface $F_{Z}$ in units of meV/{\AA}. Positive values of the
force components are directed toward an increase of $d_{\text{H-H}}$ or $Z$.}

\end{table}

\newpage

\newpage

\newpage

\end{document}